\documentclass[structabstract]{aa} 

\usepackage{graphicx}
\usepackage{amsmath}
\usepackage{txfonts}
\usepackage{natbib}

\newcommand{\bo}{\ensuremath{\boldsymbol{B}_0}}
\newcommand{\abs}[1]{\ensuremath{\left\lvert #1\right\rvert}}

\newcommand{\eps}{\ensuremath{\varepsilon}}
\newcommand{\Rl}{\ensuremath{R_{\mathrm L}}}
\newcommand{\vA}{\ensuremath{v_{\mathrm A}}}
\newcommand{\rA}{\ensuremath{R_{\mathrm A}}}

\newcommand{\be}{\begin{equation}}
\newcommand{\ee}{\end{equation}}
\newcommand{\bs}{\begin{subequations}}
\newcommand{\es}{\end{subequations}}

\newcommand{\ez}{\ensuremath{\hat{\boldsymbol{e}}_z}}

\newcommand{\pa}{\ensuremath{_\parallel}}

\newcommand{\De}{\ensuremath{\varDelta}}
\newcommand{\Om}{\ensuremath{\varOmega}}

\newcommand{\Ga}{\ensuremath{\varGamma}}
\newcommand{\La}{\ensuremath{\varLambda}}
\newcommand{\Ps}{\ensuremath{\varPsi}}

\newcommand{\uint}{\ensuremath{\int_{-\infty}^\infty}}

\newcommand{\f}[1]{\ensuremath{\boldsymbol{#1}}}
\newcommand{\m}[1]{\ensuremath{\left\langle #1\right\rangle}}

\newcommand{\pd}[2][]{\ensuremath{\frac{\partial #1}{\partial #2}}}
\newcommand{\dd}[2][]{\ensuremath{\frac{\mathrm{d} #1}{\mathrm{d} #2}}}
\newcommand{\df}{\ensuremath{\mathrm{d}}}

\begin{document}
\title{Modification of cosmic-ray energy spectra by\\stochastic acceleration}
\author{R.\,C. Tautz\inst1 \and I. Lerche\inst2 \and F. Kruse\inst1}
\institute{Zentrum f\"ur Astronomie und Astrophysik, Technische Universit\"at Berlin, Hardenbergstra\ss{}e 36, D-10623 Berlin, Germany\\
\email{rct@gmx.eu}
\and
Institut f\"ur Geowissenschaften, Naturwissenschaftliche Fakult\"at III, Martin-Luther-Universit\"at Halle, D-06099 Halle, Germany\\
\email{lercheian@yahoo.com}}

\date{Received \today; accepted April 4, 2063}

\abstract
{Typical space plasmas contain spatially and temporally variable turbulent electromagnetic fields. Understanding the transport of energetic particles and the acceleration mechanisms for charged particles is an important goal of today's astroparticle physics.}
{To understand the acceleration mechanisms at the particle source, subsequent effects have to be known. Therefore, the modification of a particle energy distribution, due to stochastic acceleration, needs to be investigated.}
{The diffusion in momentum space was investigated by using both a Monte-Carlo simulation code and by analytically solving the momentum-diffusion equation. For simplicity, the turbulence was assumed to consist of one-dimensional Alfv\'en waves.}
{Using both methods, it is shown that, on average, all particles with velocities comparable to the Alfv\'en speeds are accelerated. This influences the energy distribution by significantly increasing the energy spectral index.}
{Because of electromagnetic turbulence, a particle energy spectrum measured at Earth can drastically deviate from its initial spectrum. However, for particles with velocities significantly above the Alfv\'en speed, the effect becomes negligible.}

\authorrunning{Tautz et al.}
\titlerunning{Modification of cosmic-ray energy spectra by\\stochastic acceleration}
\maketitle

\section{Introduction}

Understanding the acceleration processes of cosmic rays has been a long-standing problem in astrophysics \citep[e.g.,][]{dru83:acc,bla87:cro}. It is currently accepted that high-energy particles with energies up to $10^{15}$\,eV are accelerated in supernova remnants \citep[e.g.,][]{abd10:acc,har83:snr} via diffusive shock acceleration, thus producing a power-law energy spectrum. The underlying processes date back to \citet{fer49:crs} and have since been improved \citep{duf05:acc} by incorporating many additional effects \citep[see][for an overview]{lon11:hea}. The resulting energy spectrum agrees both with measurements \citep[e.g.,][]{nag00:uhe,let11:uhe} and with laser experiments \citep[e.g.,][]{kur11:cra}. Particles with low energies up to $10^{10}$\,eV, instead, originate from the Sun via coronal mass ejections.

In either case, however, there are many processes that may change the energy spectrum of the particles on their way to Earth \citep{mal10:acc}. Energy-dependent loss and cooling processes, for example, modify the energy distribution function \citep{but01:sto}, generally because the interstellar (and the interplanetary) medium contains turbulent electric fields. Based on Mariner~2 measurements, it was found \citep{bel71:alf} that, to some extent, the heliospheric plasma is dominated by outward-propagating Alfv\'enic turbulence, which has been named the slab component. However, Alfv\'enic waves have also been assumed for the investigation of cosmic-ray acceleration at the galactic center \citep{fat11:acc,fat12:acc}, for example.

Now the presence of electromagnetic turbulence gives rise to reacceleration processes \citep{hei95:rea} so that particles gain additional energy while propagating through an otherwise homogeneous medium. The anomalous cosmic-ray component is also influenced by stochastic acceleration processes \citep{mor08:ano}, thereby giving rise \citep{fis07:tai,fis10:acc} to a suprathermal tail in the distribution function with $-5$ as the energy spectral index. We note, however, that thermodynamic arguments and the assumption of compressional turbulence may be sufficient to derive a similar spectral index \citep{fis07:the}. Likewise, adiabatic cooling is a more efficient process than the interaction with turbulent electric fields \citep{zha11:acc,lit11:fac,tau12:adf}. In anisotropic helical turbulence, large-scale electric fields may also give rise to particle acceleration \citep{fed08:sto}.

For these reasons, much effort has been focused on the general properties of momentum diffusion \citep[e.g.,][]{mic99:mom,lee75:mhd,ach81:fer} in slab and isotropic turbulence with Alfv\'en and oblique propagating fast magnetosonic waves, respectively. Furthermore, electrons near the Sun may also be accelerated thanks to the interaction with Whistler waves \citep{ham92:fla,voc05:whi,voc08:sup}. Momentum diffusion \citep{rs:rays,sch94:mom} can be considered as the underlying process of stochastic acceleration, where particles mostly gain significant amounts of kinetic energy. If spatial diffusion is to be neglected, stochastic acceleration can be reduced to pure momentum diffusion \citep{ost97:sof}. The connection of stochastic acceleration and momentum diffusion has been investigated analytically using magnetohydrodynamic (MHD) waves \citep[e.g.,][]{ski75:alf,sch89:cr1} by applying quasi-linear theory \citep{sha04:mhd}.

In the presence of shock waves \citep{sch89:cr2}, the first-order Fermi acceleration is usually dominant compared to stochastic acceleration in the downstream region \citep{vai99:alf}. In contrast, in this article the stochastic acceleration outside a shock wave is investigated by means of test-particle simulations. Such codes neglect the influence of the particles on the turbulence \citep{gia99:sim,mic99:mom,tau10:pad}, which is generally justified if concentration is focused on a tenuous component such as cosmic rays. Because the turbulence consists of MHD plasma waves that carry an electric field, momentum diffusion occurs so that the average particle energy is changed. In most cases, particles with discrete energies are assumed so that the energy-dependence of transport parameters such as the mean free path can be investigated. Here, in contrast, an initial distribution function is assumed, the time evolution of which is then traced so that the influence of the stochastic acceleration can be evaluated. Furthermore, the two cases will be distinguished by (i) an ensemble that evolves without supplying new particles, and (ii) continuous particle injection according to the prescribed initial distribution. Case~(i) has been applied to solar energetic particle events \citep{dro05:pro,sch09:atp}. Case~(ii) corresponds to the generation of the anomalous cosmic-ray spectrum \citep{fis06:spe}. Therefore, we intent to demonstrate that Monte-Carlo test-particle simulations are a suitable tool for tracking the time evolution of a particle velocity distribution or an energy-spectrum.

The paper is organized as follows: In Sect.~\ref{turb}, the basic properties of the electromagnetic turbulence model are presented. In Sects.~\ref{sim} and \ref{res}, the numerical test-particle code \textsc{Padian} is introduced and the simulations results are presented, respectively. In Sect.~\ref{momdiff}, an analytical solution of the momentum-diffusion equation is derived, and is compared to the numerical results. A brief summary and a discussion of the results with regard to future applications are given in Sect.~\ref{summ}.

\section{Astrophysical turbulence properties}\label{turb}

In astrophysical plasmas such as the heliosphere, turbulent electromagnetic fields are generated by instabilities or by large-scale motion cascaded towards smaller scales. Measurements in the solar wind \citep[e.g.,][]{bru05:sol} confirm the \citet{kol91:tur} power law, which gave rise to a spectral turbulent energy distribution of the form \citep[e.g.,][]{bie94:pal,tau06:wav}
\be\label{eq:spect}
G(k)\propto\ell_0\left(\delta B\right)^2\left[1+(\ell_0k)^2\right]^{-5/6},
\ee
where $\ell_0=0.03$\,a.u. denotes the turbulence bend-over scale.

In addition to the turbulence spectrum, the particle behavior is also influenced by the turbulence geometry, i.e., the angular distribution with respect to a preferred direction, which is given by an ambient magnetic field such as the solar magnetic field \citep{par58:spi,fis96:mag}. Apart from fully three-dimensional anisotropic models \citep{mat90:mal,wei10:mal,rau12:mal}, there are three models that have been used over the years: (i) the slab model, which assumes that the turbulent field depends solely on the coordinate along the mean magnetic field $\bo$ as $\delta\f B=\delta\f B(z)$ for $\bo=B_0\ez$; (ii) the composite model, which superposes the slab field by a two-dimensional complement as $\delta\f B=\delta\f B_{\text{slab}}(z)+\delta\f B_{\text{2D}}(x,y)$; and (iii) the isotropic model, which does not assume a preferred direction for the turbulent field. For reasons of simplicity, the slab model is used here.

For the propagating plasma waves, which enter Eq.~\eqref{eq:dB} via the wave frequency $\omega$, specifically linearly polarized, undamped slab Alfv\'en waves are chosen. In that case, the dispersion relation reads
\be\label{eq:disprel}
\omega(\f k)=\pm\vA k\pa,
\ee
where the Alfv\'en speed is $\vA=B_0/\!\sqrt{4\pi\rho}$ with $\rho$ the mass density. Other possible wave types such as slow and fast magnetosonic waves or Whistler waves require three-dimensional turbulence, something that is beyond the scope of the present investigation.

\section{Monte-Carlo simulations}\label{sim}

For the test-particle simulations, the test-particle Monte-Carlo code \textsc{Padian} \citep{tau10:pad} is used, which solves the trajectories of a large number of particles as they are being scattered by electromagnetic turbulence. Dimensionless variables are introduced as $\tau=\Om t$ for the time with $\Om=qB/(mc)$ the Larmor frequency, and $\f R=\gamma\f v/(\Om\ell_0)$ as a dimensionless rigidity per magnetic field with $\ell_0$ a characteristic length scale (see below).

\subsection{Numerical turbulence generation}

Numerically, the turbulent electromagnetic fields are obtained from the superposition of a number of propagating MHD plasma waves as
\be\label{eq:dB}
\delta\f B(\f r,t)=\text{Re}\sum_{n=1}^{N_m}\f e'_\perp A(k_n)\exp\!\left\{i\left[k_nz'-\omega(k_n)t+\beta_n\right]\right\},
\ee
where the wavenumbers $k_n$ are distributed logarithmically in the interval $k_{\text{min}} \leqslant k_n \leqslant k_{\text{max}}$ and where $\beta$ is a random phase angle. For the amplitude and the polarization vector, one has $A(k_n)\propto\sqrt{G(k_n)}$ and $\f e'_\perp\cdot\f e'_z=0$, respectively, with the primed coordinates determined by a rotation matrix with random angles.

With these assumptions and for the variables introduced above, the Newton-Lorentz equation can be expressed as
\be
\dd\tau\,\f R=\rA\frac{\delta B}{B_0}\:\hat{\f e}_{\hspace{0.2pt}\delta\hspace{-0.2pt}E}+\f R\times\left(\hat{\f e}_{B_0}+\frac{\delta B}{B_0}\:\hat{\f e}_{\hspace{0.2pt}\delta\hspace{-0.2pt}B}\right), \label{eq:NLe}
\ee
where $\hat{\f e}_{\hspace{0.2pt}\delta\hspace{-0.2pt}B}$ and $\hat{\f e}_{B_0}$ are unit vectors in the directions of the turbulent and the background magnetic fields, respectively. The unit vector for the electric field $\hat{\f e}_{\hspace{0.2pt}\delta\hspace{-0.2pt}E}$ is obtained using Faraday's induction law \citep{rs:rays}. Furthermore, the parameter $\rA\equiv\vA/(\ell_0\Om)$ is called the Alfv\'en rigidity.

\subsection{Diffusion coefficients}

In the normalized coordinates, spatial diffusion coefficients are defined through the mean square displacements in the directions parallel and perpendicular to the ambient magnetic field as \citep{tau10:pad}
\be
\frac{\kappa_{\parallel,\perp}(t)}{\ell_0^2\Om}=\frac{1}{2\Om t}\m{\left[\De x_{\parallel,\perp}(t)\right]^2}.
\ee
Here, $\kappa$ is time-dependent and hence is frequently denoted the running diffusion coefficient.

Using the same general derivation, a dimensionless running momentum diffusion coefficient can be defined through \citep[cf.][]{ost93:mom,tau10:wav}
\be\label{eq:Dp}
\frac{D_p(p,t)}{p_{\text{init}}^2\Om}=\frac{1}{R_{\text{init}}^2\Om t}\m{\left[\f R(t)-\f R_{\text{init}}\right]^2}.
\ee
From the inverse momentum diffusion coefficient, the so-called acceleration time scale \citep[e.g.,][]{zan06:sho,osu09:lob,dos10:sho} can be constructed as
\be
\tau_{\text{acc}}\equiv\frac{p}{\dot p}\approx\frac{p^2}{D_p(p)}.
\ee
Whether or not the momentum diffusion coefficient is truly diffusive, i.e., whether it has a finite constant value for large times, is not entirely clear at present \citep[cf.][]{tau10:wav,tau10:sub}. Generally, the larger the ratio $\rA/R$, the sooner $D_p$ attains an asymptotic value.\footnote{At first glance is might seem puzzling that diffusion coefficients are not constant even though there is turbulence from the start. A long time ago, when comparing random particle motions with diffusive processes, \citet{cha43:sto} found that truly diffusion-like behavior the corresponding equivalent diffusion coefficient is time-dependent and reaches an asymptotic constant value only later in time.}

\subsection{Rigidity distribution}

In most test-particle simulations, a number of discrete initial particle rigidities are assumed which are then investigated separately, which is acceptable when one attempts to investigate the rigidity-dependence of the diffusion coefficients. Here, however, the goal is to trace the evolution of a given rigidity distribution function, while the particles are scattered in momentum space.

\begin{figure}[tb]
\centering
\includegraphics[width=\linewidth]{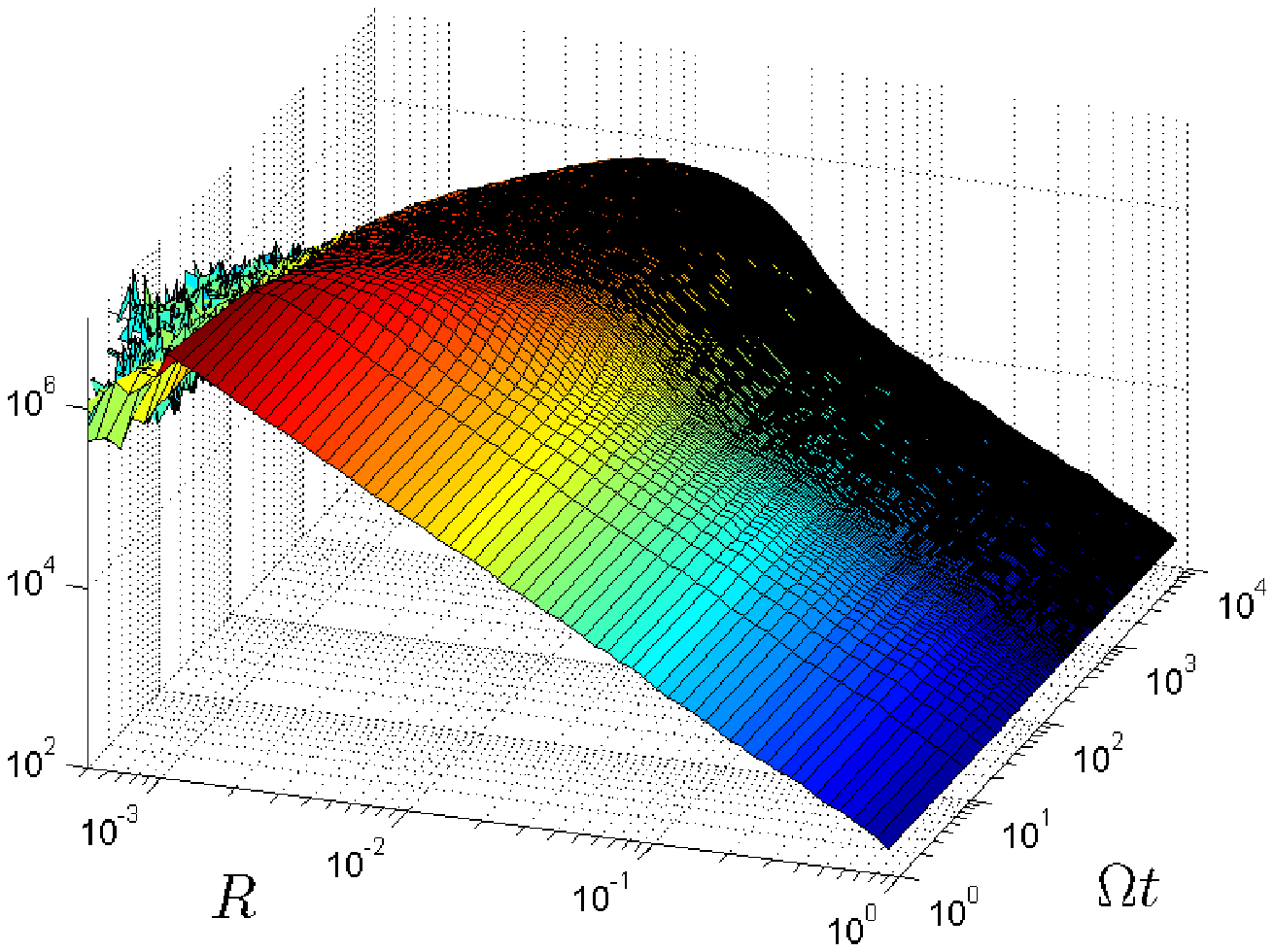}\\
\includegraphics[width=\linewidth]{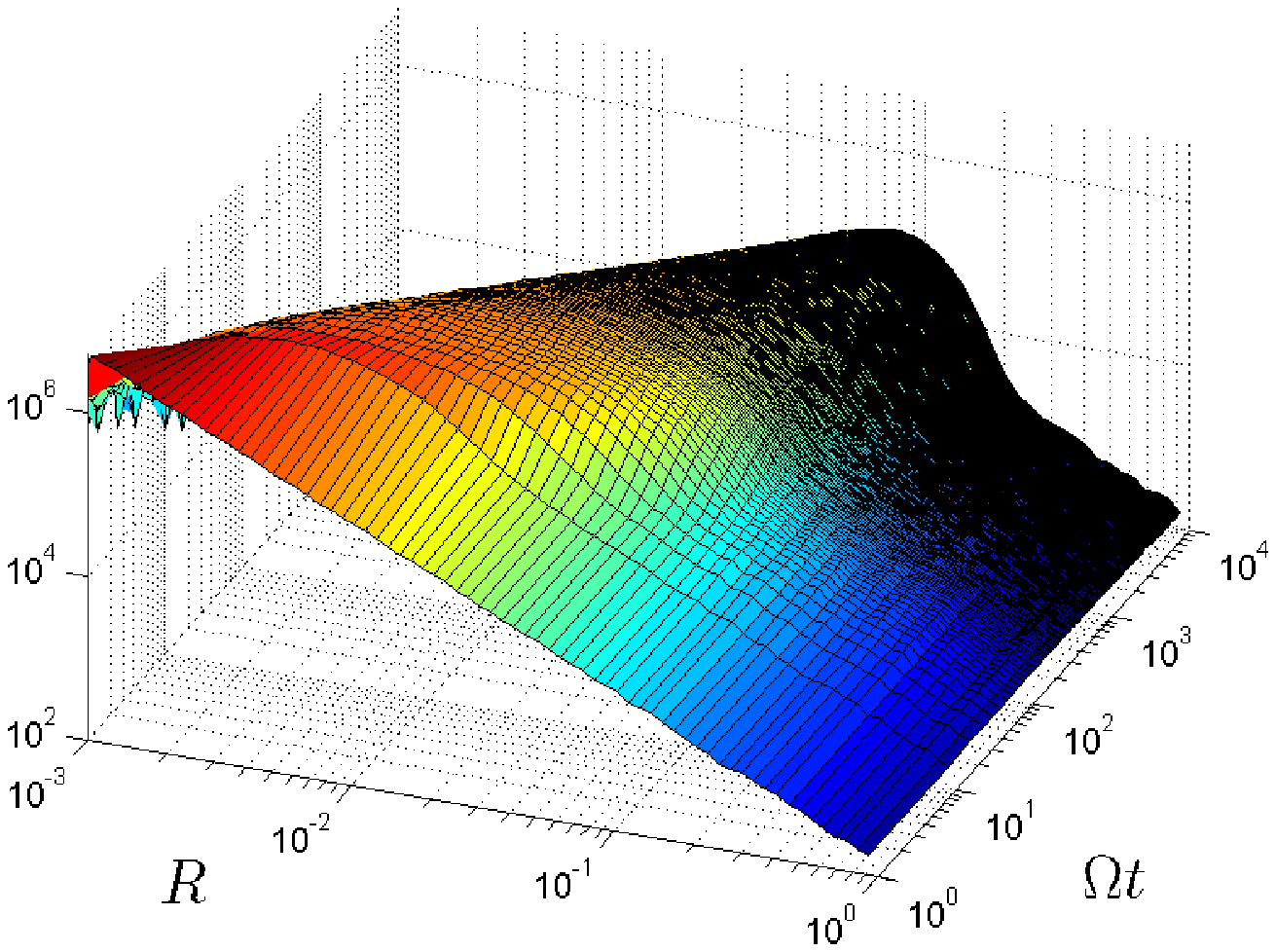}
\caption{(Color online) The evolution of the particle distribution as a function of the rigidity $R$ as the normalized time $\Om t$ increases. For the Alfv\'en rigidity, values of $\rA=10^{-4}$ (upper panel) and $\rA=10^{-3}$ (lower panel) have been used.}
\label{ab:distri}
\end{figure}

In accordance with cosmic-ray observations \citep[see][for recent observations]{let11:uhe}, the initial rigidity distribution is modeled in the form of a power law. The requirement that the distribution function be normalized to unity leads to
\bs\label{eq:dist}
\be
f(R,a)=\frac{1-a}{R_{\text{max}}^{1-a}-R_{\text{min}}^{1-a}}\;R^{-a}
\ee
in $R_{\text{min}}\leqslant R\leqslant R_{\text{max}}$ and $f(R,a)=0$ otherwise. We note that Eq.~\eqref{eq:dist} is valid for $a\neq1$, whereas for $a=1$ one has the special form
\be
f(R,1)=\left(\ln\frac{R_{\text{min}}}{R_{\text{max}}}\right)^{-1}\frac{1}{R}
\ee
\es
again in $R_{\text{min}}\leqslant R\leqslant R_{\text{max}}$. For the numerical simulations, the spectral index has to be chosen so that, on the one hand, a relatively steep initial energy spectrum is obtained. On the other hand, the steeper the spectrum, the fewer particles will be found at the high-energy end of the spectrum, thus requiring a larger total number of particles. As a compromise, here the value $a=1.5$ will be used. The same holds true for the minimum and maximum rigidity values. Whereas the interpretation of the numerical simulations requires a large number of particles per rigidity interval, $R_{\text{min}}$ and $R_{\text{min}}$ are chosen so that particles with rigidities comparable to and much larger than $\rA$ are present but, at the same time, the rigidity interval is kept moderately narrow.

\begin{figure}[tb]
\centering
\includegraphics[width=\linewidth]{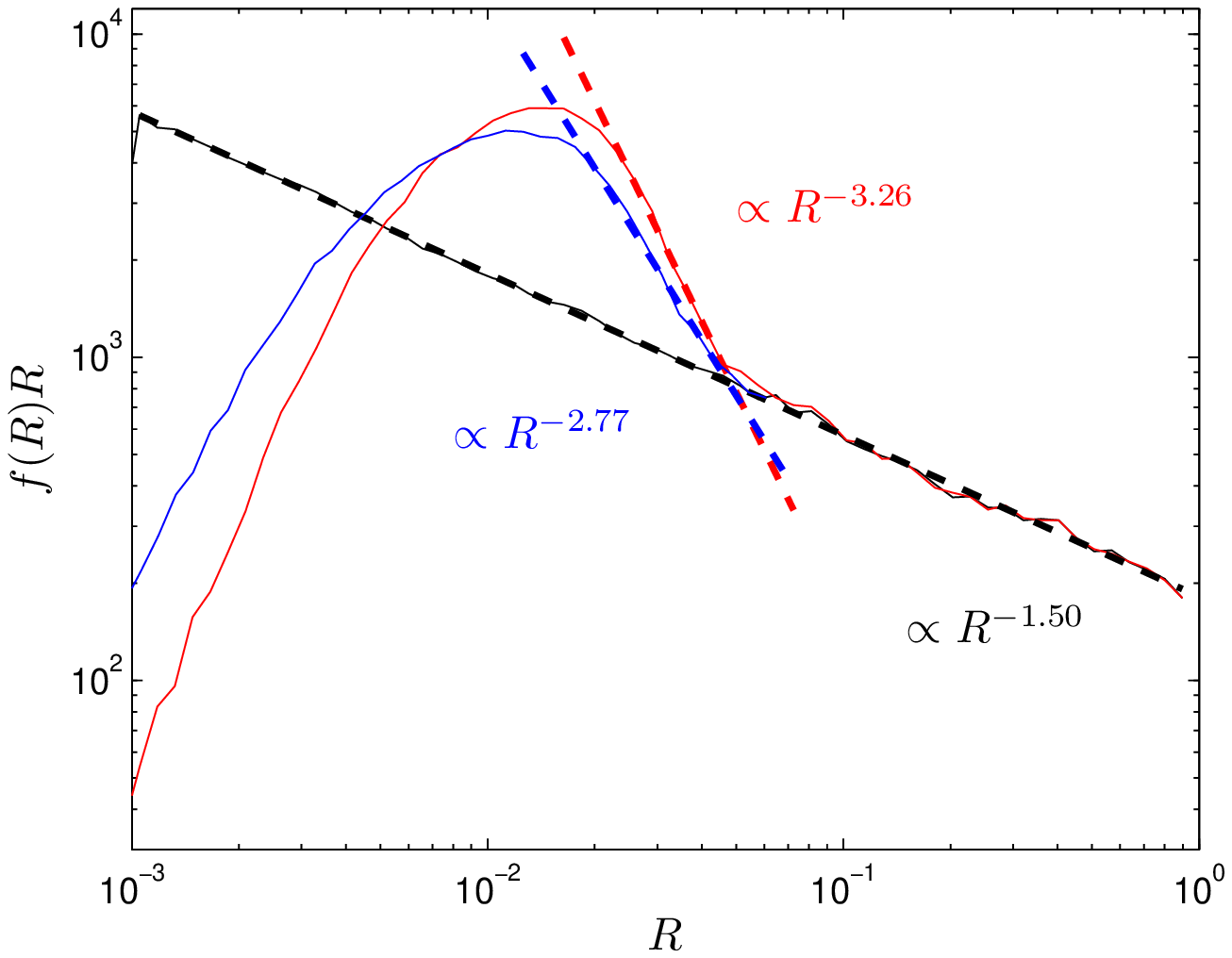}\\
\includegraphics[width=\linewidth]{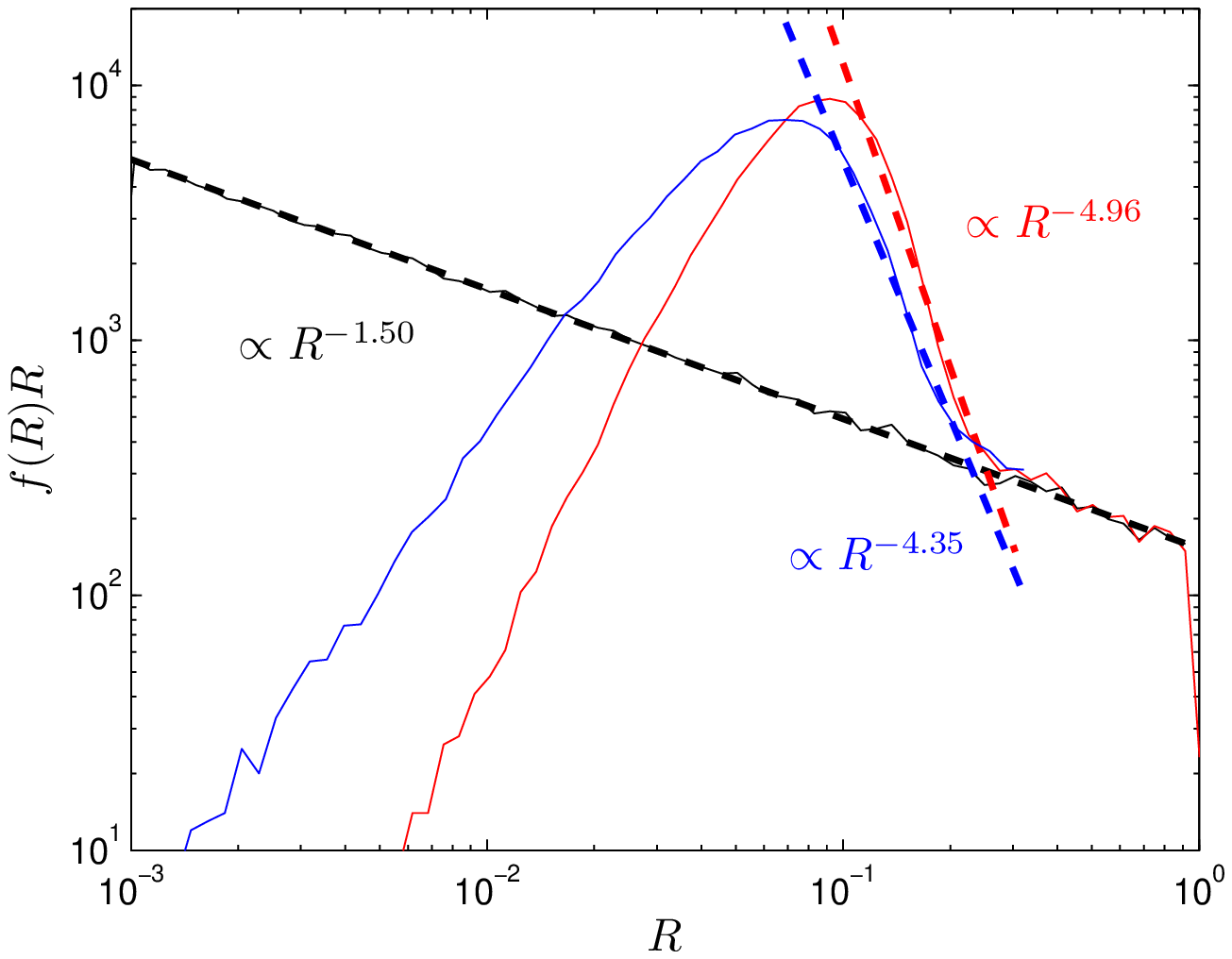}\\
\caption{(Color online) The energy spectral index for the cases of $\rA=10^{-4}$ (upper panel) and $\rA=10^{-3}$ (lower panel). Shown is the energy distribution function at the beginning (black lines) and at the end of the simulation for the cases of single injection (red lines) and continuous particle injection (blue lines). Also shown are the spectral indices obtained from power-law fits (dashed lines) to the simulation data (thin lines).}
\label{ab:index}
\end{figure}

In the simulation code, particle rigidities are randomly drawn from the distribution function (Eq.~\eqref{eq:dist}) using the transformation method \citep[e.g.,][p.~362]{pre07:nr3}. The method requires uniform random numbers $x\in[0,1]$ to be fed to the inverse cumulant $F$ of the distribution function, which reads
\bs
\be
F^{-1}(x,a)=\left[R_{\text{min}}^{1-a}+x\left(R_{\text{max}}^{1-a}-R_{\text{min}}^{1-a}\right)\right]^{1/(1-a)}
\ee
for $a\neq1$ and
\be
F^{-1}(x,1)=R_{\text{max}}^xR_{\text{min}}^{1-x}
\ee
\es
if $a=1$, thereby returning rigidity values $R$ with the probability prescribed by Eq.~\eqref{eq:dist}.

\section{Simulation results}\label{res}

Two basic sets of simulations need to be distinguished: (i) single injection, where all particles are injected at time $t=0$ and their trajectories are monitored, and (ii) continuous injection, where, throughout the simulation, new particles are injected with rigidities determined according to the distribution function from Eq.~\eqref{eq:dist}. Technically, case~(ii) is realized by choosing the maximum simulation time as a random number so that, throughout the simulation, particles with random life times are present. All simulation parameters are summarized in Table~\ref{ta:param}.

\begin{figure}[tb]
\centering
\includegraphics[width=\linewidth]{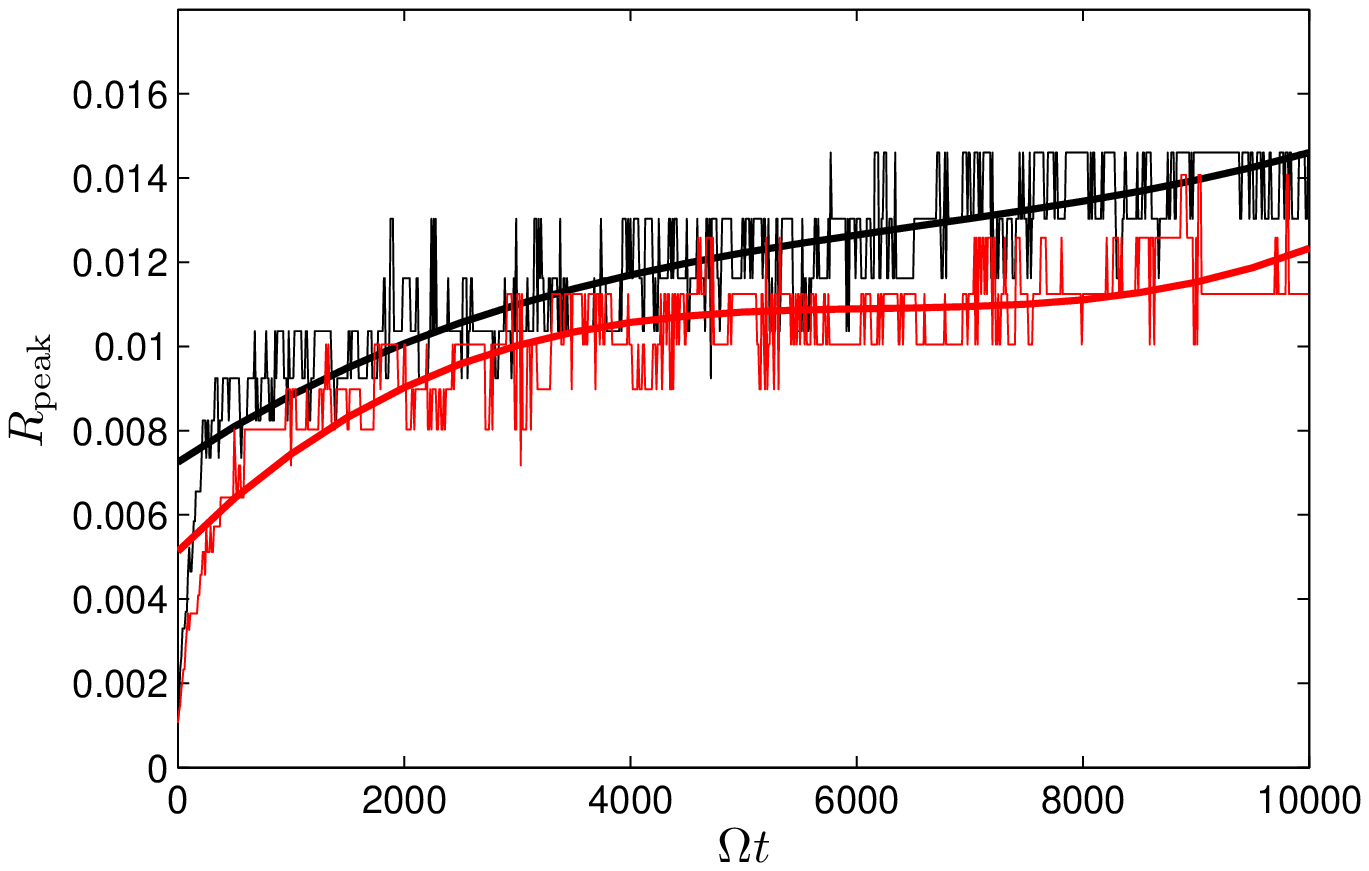}\\
\includegraphics[width=\linewidth]{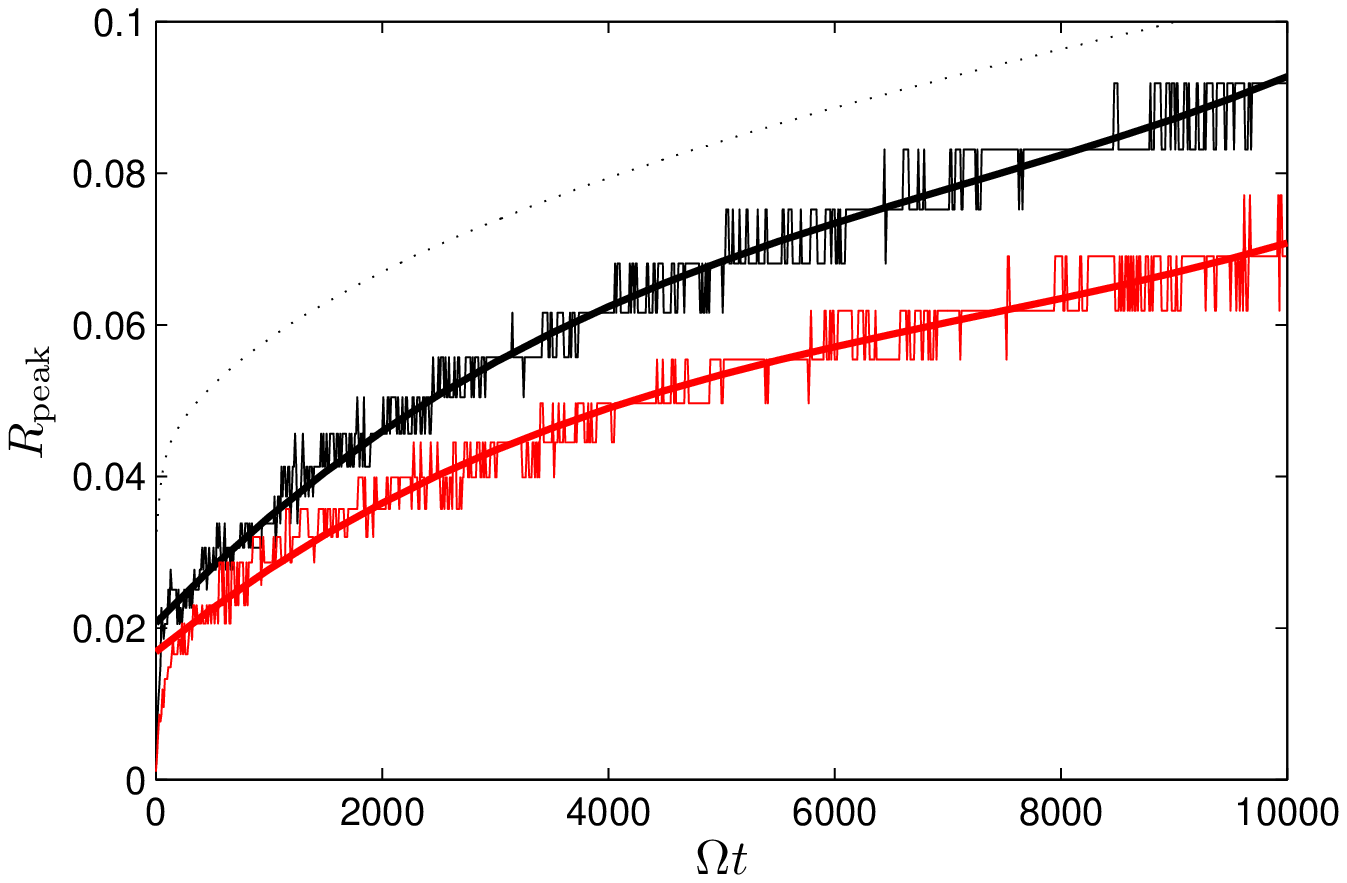}
\caption{(Color online) The maximum of the distribution function as the time increases. Again the cases of $\rA=10^{-4}$ (upper panel) and $\rA=10^{-3}$ (lower panel) are shown. The black and red lines denote the cases of single injection and continuous particle injection, respectively. The thick lines simply show cubic fits of the actual simulation data. Additionally, the \emph{average} particle rigidity is shown in the lower panel (dotted line).}
\label{ab:Rmax}
\end{figure}

\begin{table}[b]
\centering
\begin{tabular}{lll}\hline\hline
Parameter Description			& Symbol			& Value\\\hline
Minimum rigidity				& $R_{\text{min}}$	& $10^{-3}$\\
Maximum rigidity				& $R_{\text{max}}$	& $1$\\
Alfv\'en rigidity				& $\rA$			& $10^{-4;\;-3}$\\
Energy spectral index			& $a$			& $1.5$\\
Number of plane waves			& $N$			& $256$\\
Number of particles				& ---			& $10^5$\\
Relative turbulence strength		& $\delta B/B_0$	& $1$\\
\hline\hline
\end{tabular}
\caption{Parameters used for the Monte-Carlo simulation. The number of particles used for the averaging process is subdivided into 250~different turbulence realizations corresponding to different sets of random numbers, in each of which the trajectories of 400~particles are calculated.}
\label{ta:param}
\end{table}

\begin{figure}[tb]
\centering
\includegraphics[width=\linewidth]{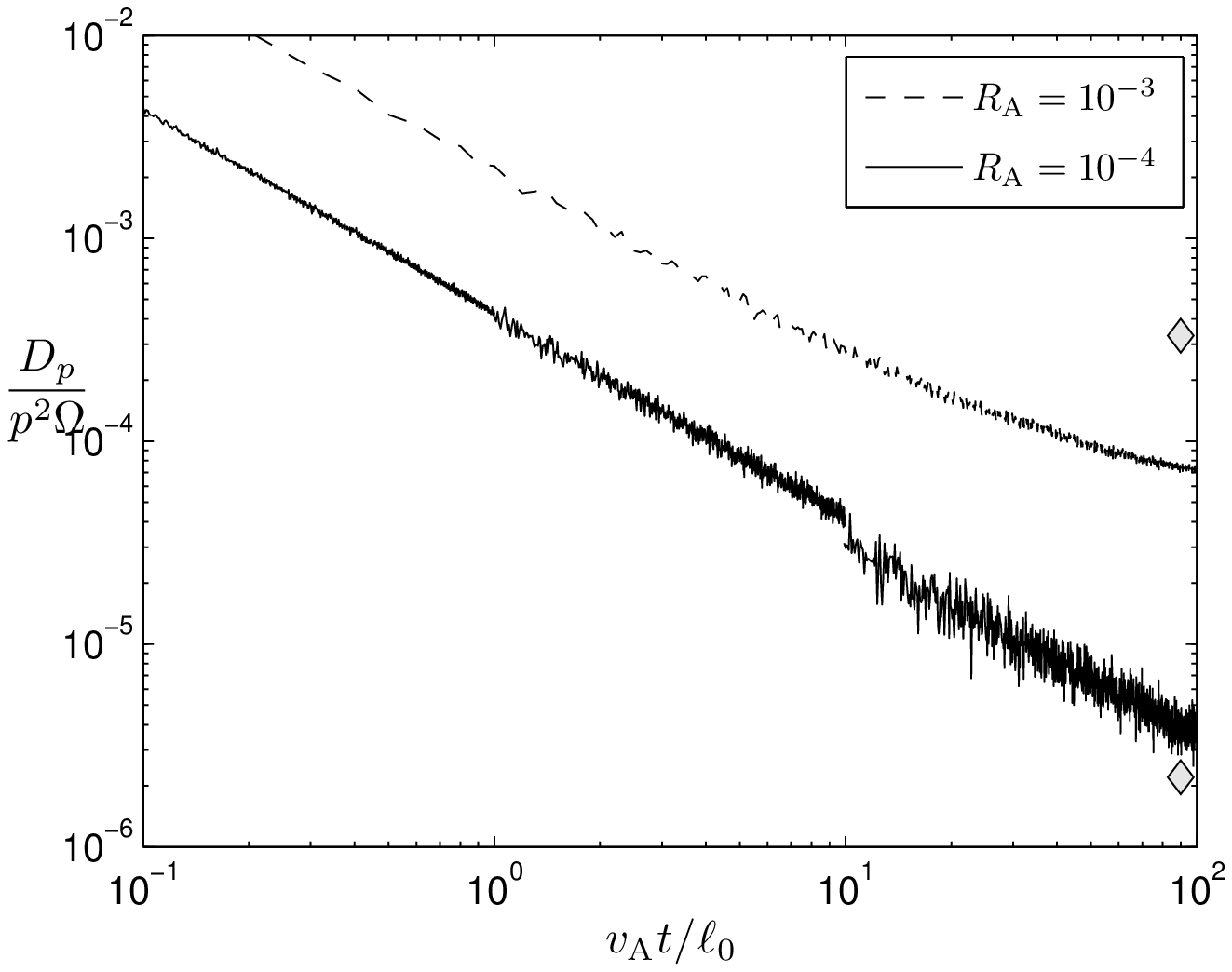}
\caption{The numerically calculated running momentum diffusion coefficient for the cases of $\rA=10^{-4}$ (solid line) and $\rA=10^{-3}$ (dashed line). We note that for the last part of the lower curve, only $10^3$ particles could be used, while the rest was obtained from a simulation with $10^4$ particles. The two diamonds illustrate the quasi-linear result from Eq.~\eqref{eq:DpQLT} in Appendix~\ref{app:momdiff}.}
\label{ab:TmomMH}
\end{figure}

In Fig.~\ref{ab:distri}, the evolution of the distribution function is traced as time increases. Initially, a power-law distribution according to Eq.~\eqref{eq:dist} is assumed. As the simulation time increases, a maximum in the distribution function is quickly established. Because of the stochastic acceleration of the particles, the maximum is then constantly shifted towards higher energies. In the case of a low Alfv\'en velocity, the process proceeds relatively slowly. We note that the results shown in Fig.~\ref{ab:distri} correspond to case~(i).

The quantitative variation of the energy spectral index is illustrated in Fig.~\ref{ab:index}, which illustrates the time evolution of the velocity distribution by comparing the initial and the final particle spectra. Whereas initially all particles are distributed according to the power law in Eq.~\eqref{eq:dist}, this is later changed towards a broken power-law distribution (upper panel) and, for larger Alfv\'en velocities, a distribution that shows a power-law feature only in a very narrow energy range. In all simulation runs, the spectral indices have been fitted to the steepest range. The notable result is that in the case of continuous injection the spectrum is only moderately flattened even though newly injected particles are present at all times.

The maximum particle rigidity is depicted in Fig.~\ref{ab:Rmax}, showing a steady increase due to stochastic acceleration. The only exception is found for low Alfv\'en rigidity in the case of continuous injection, where the average rigidity remains constant for some time. Again, it is the Alfv\'en rigidity that has the prominent influence on the resulting average particle energy, at least qualitatively, and not the average particle life-time as distinguished by continuous versus single injection.

The associated running momentum diffusion coefficient determined through Eq.~\eqref{eq:Dp} is shown in Fig.~\ref{ab:TmomMH}. We note that the time is normalized as $\vA t/\ell_0$, which illustrates that for smaller Alfv\'en velocities larger times are needed until the momentum diffusion coefficient reaches its asymptotic limit. It is precisely the fact that almost a $1/t$ time dependence can be seen before finally a transition to a constant (diffusive) value occurs that makes it so numerically difficult to obtain a reliable value for the momentum diffusion coefficient. The maximum simulation time $\vA t/\ell_0=10^2$ corresponds to $\Om t=10^5$ and $\Om t=10^6$, respectively, which illustrates that for the parameters chosen here, scattering in momentum takes an extremely long time until the diffusive limit is reached.\footnote{Of course, ``long'' is a relative term as $10^6$ gyro periods is a short time compared to the dynamical time scales in most astronomical sources. However, both computationally and with respect to spatial diffusion---which takes about a factor of $10^4$ fewer gyro periods---that time scale is indeed extremely long.}

\section{Analytical stochastic acceleration}\label{momdiff}

In principle, the stochastic variations in the particle energy may be positive or negative. However, as shown in Sect.~\ref{res}, the net effect was a significant energy gain of the particle ensemble. To verify this finding, we present an approach to this problem by directly solving the equation of momentum diffusion. The behavior of charged particles in turbulent electromagnetic fields has to be described using a Fokker-Planck approach \citep{par65:pas,rs:rays} or at least a diffusion-convection equation. Usually, the terms that describe spatial diffusion are of higher magnitude than momentum-diffusion terms. For particular parameter values analytical solutions are feasible \citep[e.g.,][]{bus05:snr,poh09:crd}, whereas a full solution requires numerical tools \citep{guy05:pde}.

However, if concentration is to be focused on stochastic energy changes only \citep[e.g., Sect.~3.3 in][]{tve67:fer,bla87:cro,ost97:sof}, it is sometimes convenient to neglect spatial diffusion altogether. Under certain conditions\footnote{For example, the principle of detailed balance has to be fulfilled, which requires an equilibrium state \citep[e.g.,][]{kle55:bal}. Specifically, in the present case it states that the probability $\Ps(\f p,\f p')$ for changing the particle momentum from $\f p$ to $\f p'$ satisfies the condition $\Ps(\f p,\f p')=\Ps(\f p',\f p)$.} the diffusion-convection equation can then be expressed as
\be\label{eq:momdif}
\pd[f]t=\frac{1}{p^2}\,\pd p\left(p^2D_p(p)\,\pd[f]p\right),
\ee
which is a linear parabolic partial differential equation. In principle, $D_p$ depends also on time, which however will be neglected here for reasons of simplicity. We note that, by defining a dimensionless rigidity diffusion coefficient $D_R=D_p/(\ell_0^2m^2\Om^3)$, Eq.~\eqref{eq:momdif} can be written in dimensionless form by using the normalized variables introduced in Sect.~\ref{sim}.

For Alfv\'en waves, the diffusion coefficient \citep[see, e.g.,][]{sch85:dis,sch89:cr1,sta08:mom} reads
\be\label{eq:DpA}
D_p(p)\approx\left(\frac{\delta B}{B_0}\right)^2\left(\frac{\vA}{c}\right)^2\frac{p^2c}{\Rl^{2-s}}\left(\frac{2\pi}{k_{\text{max}}}\right)^{1-s}
\ee
with $\Rl$ the Larmor radius, so that $D_p\propto p^s$. It should be noted however that Eq.~\eqref{eq:DpA} was derived only for a simple turbulence spectrum of the form $G(k)\propto k^{-s}$ with $s\in[1,2]$ (for a more detailed discussion see, e.g., \citealt[Sect.~13.1.3]{rs:rays}).

Consider the diffusion coefficient written as $D_p=D_0p^q$ and the initial distribution written in the form $f(p,a)=f_0p^{-a}$. Then Eq.~\eqref{eq:momdif} is given as
\be\label{eq:momdif2}
\pd[f]t=\frac{D_0}{p^2}\,\pd p\left(p^{q+2}\,\pd[f]p\right).
\ee
In general, there are two options. First, when $f$ is initially prescribed as the power law in Eq.~\eqref{eq:dist}, then a Fourier approach can be used as shown in Appendix~\ref{app:fourier}. This allows the reduction of the momentum diffusion equation to an ordinary differential equation but has the disadvantage that the Fourier transform of the initial distribution function must be known. Second, when $f$ is initially unknown, a self-similar solution can be sought, which is based on the separation $f(t,R)=S(t)H\bigl(p/E(t)\bigr)$ and will be discussed in the next subsection.

\subsection{Self-similar solution}

Consider again Eq.~\eqref{eq:momdif2}. For a self-similar solution one seeks a solution in the form
\be\label{eq:fsim}
f(t,p)=S(t)H\left(\frac{p}{E(t)}\right),
\ee
where $S$, $H$, and $E$ need to be determined. With $\zeta=p/E(t)$, Eq.~\eqref{eq:momdif2} can then be written
\be
\frac{1}{E^{2-q}}\left.\pd[f]t\right\rvert_\zeta+\zeta\,\pd[f]\zeta-\frac{\dot E}{E^{3-q}}=\frac{D_0}{\zeta^2}\,\pd\zeta\left(\zeta^{q+2}\,\pd[f]\zeta\right)
\ee
which, by using Eq.~\eqref{eq:fsim}, yields
\begin{equation*}
H\,\frac{\dot S}{E^{2-q}}-S\zeta\,\frac{\dot E}{E^{3-q}}\,\dd[H]\zeta=\frac{D_0S}{\zeta^2}\,\dd\zeta\left(\zeta^{q+2}\,\dd[H]\zeta\right).
\end{equation*}

\subsubsection{Solution for general $q\neq2$}

Collecting terms, one has
\be
\frac{\dot S}{E^{2-q}S}-\frac{\dot E}{E^{3-q}}\,\frac{\zeta}{H}\,\dd[H]\zeta=\frac{D_0}{\zeta^2H}\,\dd\zeta\left(\zeta^{q+2}\,\dd[H]\zeta\right),
\ee
which is separable when (the coefficients $C_1$, $C_2$, and $C_3$ are constants)
\bs
\begin{align}
\frac{\dot E}{E^{3-q}}&=C_1\\
\dd[E^{q-2}]t&=-C_1\left(2-q\right)
\intertext{so that}
E&=\left[C_2-C_1\left(2-q\right)t\right]^{1/(2-q)}.
\end{align}
Then additionally one must have
\be
\frac{\dot S}{E^{2-q}S}=\frac{\dot S}{S}\left[C_2-C_1\left(2-q\right)t\right]\equiv C_3,
\ee
\es
which integrates trivially so that one obtains $S(t)$ in the form
\be
S(t)=S_0\left[C_2-C_1\left(2-q\right)t\right]^{-C_3/[C_2(2-q)]}
\ee
and one has
\be
\dd\zeta\left(\zeta^{q+2}\,\dd[H]\zeta\right)+\frac{C_1}{D_0}\,\zeta^3\dd[H]\zeta-\frac{C_3}{D_0}\,\zeta^2H=0.
\ee

For further simplification use the fact that the distribution function is normalized, i.e.,
\be
1=S(t)E(t)^3\int_0^\infty\df\zeta\;\zeta^2H(\zeta)\equiv S(t)E(t)^3H_0
\ee
so that $S(t)^{-1}=E(t)^3H_0$, yielding $\dot S/S=-3\dot E/E$ and
\begin{equation*}
\frac{\dot S}{E^{2-q}S}=-\frac{\dot E}{E^{3-q}}=C_3=-3C_1.
\end{equation*}
Finally, the resulting differential equation for $H(t)$ reads
\be
\dd\zeta\left(\zeta^{q+2}\,\dd[H]\zeta\right)+\frac{C_1}{D_0}\,\zeta^2\,\dd[(\zeta H)]\zeta=0,
\ee
which needs a numerical solution except for special values of $q$ such as $q=2$.

\subsubsection{Solution for $q=2$}

For instance when $q=2$, set $H=\zeta^\La$ so that
\be\label{eq:La}
\La^2+\La\left(3+\frac{C_1}{D_0}\right)-\frac{C_3}{D_0}=0,
\ee
yielding
\be
2\La_\pm=-\left(3+\frac{C_1}{D_0}\right)\pm\sqrt{\left(3+\frac{C_1}{D_0}\right)^2+\frac{4C_3}{D_0}}
\ee
Now at $t=0$, the distribution function $f$ is discontinuous; therefore, let $H=\zeta^{-a}$ in $\zeta_{\text{min}}\leqslant\zeta\leqslant\zeta_{\text{max}}$ and $H=0$ otherwise, where $\zeta_{\text{min}}$ and $\zeta_{\text{max}}$ are fixed values to be related eventually to $p_{\text{min}}$ and $p_{\text{max}}$.

\begin{figure}[tb]
\centering
\includegraphics[width=\linewidth]{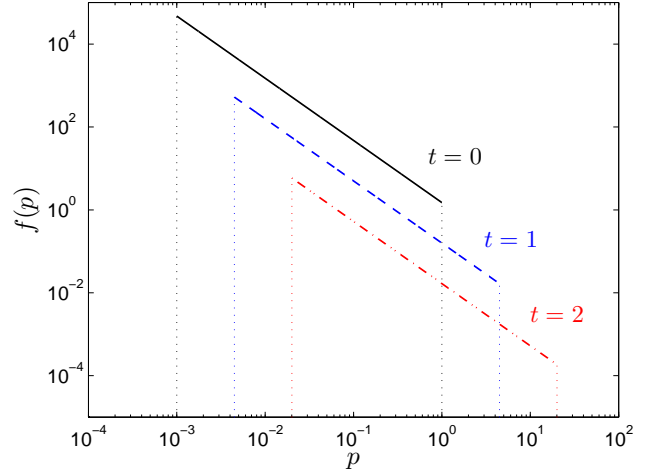}
\caption{(Color online) Evolution of the momentum distribution function as obtained from the self-similar solution of the momentum diffusion equation for the case $q=2$. The three lines illustrate the distribution at the times $t=0$ (black solid line), $t=1$ (blue dashed line), and $t=2$ (red dotdashed line).}
\label{ab:selfsimilar}
\end{figure}

The normalization requires
\bs
\begin{align}
1&=\int_0^\infty\df p\;p^2f(p)=S\int_0^\infty\df p\;p^2H\!\left(\frac{p}{E(t)}\right)\\
&=S(t)E(t)^3\int_{\zeta_{\text{min}}}^{\zeta_{\text{max}}}\df\zeta\;\zeta^2H(\zeta)\\
&=\frac{S(t)E(t)^3}{3-a}\left[(\zeta_{\text{max}})^{3-a}-(\zeta_{\text{min}})^{3-a}\right]. \label{eq:norm}
\end{align}
\es
Hence, $S(t)E(t)^3$ must be constant, which again relates the coefficients $C_3=-3C_1$. Then from Eq.~\eqref{eq:La} with $\La=-a$ one has $C_1=D_0a$.

The solution for the distribution function is then
\bs
\begin{align}
f(t,p)&=S_0E_0^a\exp\left(D_0a^2t\right)\exp\left(-3a tD_0\right)p^{-a}\\
&=S_0E_0^ap^{-a}\exp\left[-D_0a\left(3-a\right)t\right]
\end{align}
\es
in $\zeta_{\text{min}}\exp(D_0at)\leqslant p\leqslant\zeta_{\text{max}}\exp(D_0at)$. From the values at $t=0$ one therefore has $\zeta_{\text{min,max}}=p_{\text{min,max}}(t=0)$.

From the normalization condition (Eq.~\eqref{eq:norm}) one can choose $E_0=1$ so that
\be
S_0=\frac{3-a}{(\zeta_{\text{max}})^{3-a}-(\zeta_{\text{min}})^{3-a}}.
\ee
Then the distribution function $f(t,p)$ has the same normalization for all times (as must be the case); however, the minimum and maximum momenta increase as time increases. In Fig.~\ref{ab:selfsimilar}, the distribution function is shown for different times, illustrating the shift of the particles toward higher momenta.

We note that Eq.~\eqref{eq:momdif2} shows that for $q=2$ there is scale independence of the momentum (as can easily be seen by replacing $p$ by $Lp$ where $L$ is an arbitrary constant). Equation~\eqref{eq:momdif2} is unaltered by such a replacement. Thus if the initial distribution has no scale, as with a power law, then the distribution never develops a scale and retains the power law. If, however, the initial distribution is terminated at specific values of momenta then these end values must adjust with time to maintain the scale independence and also to maintain a constant particle population. This anomalous result happens only for $q=2$; for any other value of $q$ there is an overall scale for the momentum in Eq.~\eqref{eq:momdif2} so that the particle evolution with time will end up not only with adjusted end point values but also with a change in the spectral index, as indeed is shown by the numerical illustrations. Furthermore, we note that the case $q=2$ agrees with the so-called hard-sphere scattering \citep[e.g.,][]{par58:har,par95:gre} and thus possesses a tangible physical interpretation.

\begin{figure}[tb]
\centering
\includegraphics[width=\linewidth]{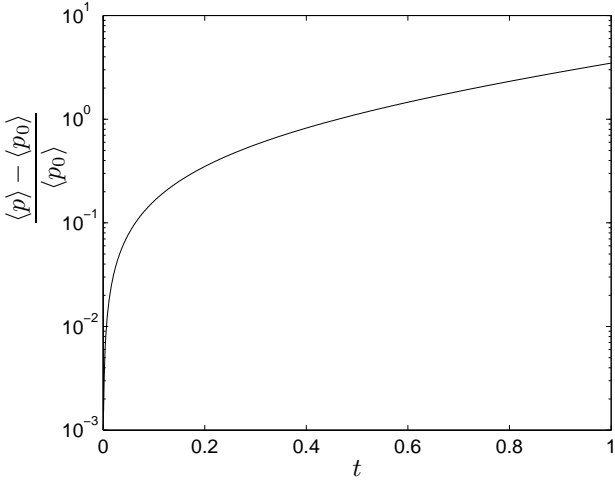}
\caption{The relative average momentum gain as described through Eq.~\eqref{eq:pave} as a function of time. The reference value is the initial average momentum, i.e., $\m{p_0}=\m{p(t=0)}$.}
\label{ab:pincrease}
\end{figure}

Finally, the average momentum gain per time can be obtained using
\be
\dd t\,\ln\m{p(t)}=D_0a,
\ee
where
\bs\label{eq:pave}
\begin{align}
\m{p(t)}&=\int_0^\infty\df p\;p^3f(t,p)\\
&=\frac{(p_{\text{max}})^{D_0a}(p_{\text{min}})^4-(p_{\text{max}})^4(p_{\text{min}})^{D_0a}}{(p_{\text{max}})^{D_0a}(p_{\text{min}})^3-(p_{\text{max}})^3(p_{\text{min}})^{D_0a}}\nonumber\\
&\times\frac{3-D_0a}{4-D_0a}\;\text e^{D_0at}.
\end{align}
\es
The relative momentum change is shown in Fig.~\ref{ab:pincrease}, thereby illustrating that particles are efficiently accelerated in agreement with the simulation results shown in Sect.~\ref{res}. Therefore, it has been confirmed that, on average, particles indeed gain energy, thus confirming the numerical ansatz.

\section{Discussion and conclusion}\label{summ}

In this paper, the stochastic acceleration of energetic charged particles, due to momentum diffusion in Alfv\'enic electromagnetic turbulence, has been investigated. The initial particle distribution function was prescribed as a power law starting with the Alfv\'en velocity and covering three orders of magnitude. Even though it was found \citep{sch89:cr1} that the dominant contribution to the stochastic acceleration of charged particles is provided by fast magnetosonic waves and not by Alfv\'en waves, here the case of Alfv\'en waves was chosen because these waves constitute the dominant time-dependent turbulent component in the solar wind. Throughout, radiative losses \citep{sch07:coo,sta08:mom,usr09:rad} have been neglected as only wave-particle interactions were included via the Lorentz force \citep[cf.][]{tau10:pad}.

Two independent methods were used:

\begin{itemize}
\item A Monte-Carlo simulation, which, in the presence of artificial turbulence obtained from the superposition of slab Alfv\'en waves, integrates the Newton-Lorentz equation for a large number of test particles. Additionally, the cases were distinguished of all particles being injected (i) at the initial time $t=0$, or (ii) continuously throughout the simulation with probabilities given by the initial distribution function.
\item By neglecting spatial diffusion, the momentum-diffusion equation has been analyzed using a self-similar approach. The solution allows one to predict the time evolution of the distribution function. A general method to obtain a solution by Fourier transforming the momentum-diffusion equation is presented in Appendix~\ref{app:fourier}.
\end{itemize}

By monitoring the average momentum value, it was confirmed in both cases that particles are efficiently \emph{accelerated}. Furthermore, the simulation results show that, for a distribution function that is initially of the form $f\propto p^{-a}$, the spectral index $a$ is significantly increased for later times. The steepened spectrum is in remarkable agreement with $p^{-5}$ spectra observed in various scenarios such as pick-up ion distributions \citep{fah11:piu}, termination-shock particles \citep{jok10:com}, and the solar corona (quiet solar wind particles) \citep{fis08:tai,ant13:sup}. We note that, because steep spectra take considerably more computing time, a relatively flat initial spectrum has been chosen; however, the agreement with observations is of itself no guarantee that this agreement cannot also be achieved with a steep injected spectrum. Demonstrating the validity of this point is, however, best delayed to a future communication. In addition, here the acceleration is due to homogeneous turbulent electromagnetic fields generated by (unstable) plasma waves. This steepening of the spectrum could not be reproduced in the analytical self-similar solution based on a momentum diffusion coefficient of the form $D_p\propto p^q$ with $q=2$ for reasons given in the text. This spectral index change requires a departure from a scale-independent momentum behavior for Eq.~\eqref{eq:momdif2}, something that occurs for all values of $q$ except $q=2$. Therefore, the latter case has been chosen for reasons of simplicity.

Furthermore, the results depend on the details of the momentum-diffusion coefficient $D_p$ for which a simplified form is assumed. As detailed in Appendix~\ref{app:momdiff}, a more general, non-linear approach shows that $D_p$ exhibits a more complicated dependence on the Alfv\'en velocity and on the turbulence power spectrum. In addition, the cases of single and continuous particle injection have been considered in the simulation. Whereas the second case might be more realistic, the first case is useful in that, by superposition of different time injection solutions, a continuous injection model can be constructed where the source term varies with time. This, however, is beyond the scope of the present investigation.

Future work should attempt to modify the analytical self-similar solution so that the effect of a variable energy spectral index is incorporated as predicted by the numerical simulations. Generally, a much larger number of test particles than is currently available is required if the evolution of a distribution function is to be investigated for a broader momentum range and/or for a considerably steeper initial spectrum. This is especially true when the particles are continuously injected so that the evolution of a quasi-equilibrium can be monitored. Furthermore, real physical scenarios such as particle acceleration at the outer regions of the solar system require the inclusion of spatial structures, thereby giving rise to spatially variable diffusion coefficients. The investigation of this problem will be deferred to a future paper.

\begin{acknowledgements}
FK thanks D. Breitschwerdt for support during his M.\,Sc. thesis. We are also appreciative of comments by a referee that helped us sharpen some otherwise fuzzy points.
\end{acknowledgements}

\appendix
\section{Fourier solution of the momentum-diffusion equation}\label{app:fourier}

The analytical solution of Eq.~\eqref{eq:momdif} for the initial distribution from Eq.~\eqref{eq:dist} proceeds as follows. There are two alternatives: (i) a direct solution via Fourier transformation; and (ii) a self-similar solution approach. While the second approach is discussed in the text, here some basic considerations are shown regarding the first approach.

Set $\df\xi=p^{-(q+2)}\df p$ so that $\xi=p^{-(q+1)}/(q+1)$ when Eq.~\eqref{eq:momdif2} takes the form
\be\label{eq:momdif3}
\pd[f]t=\frac{D_0}{p^{q+4}}\,\pd[^2f]{\xi^2}.
\ee

With the Fourier transform dependence in time proportional to $\text e^{-i\omega t}$, Eq.~\eqref{eq:momdif3} yields
\be\label{eq:momdif4}
-\frac{i\omega}{D_0}\,p^{q+4}F=\pd[^2F]{\xi^2},
\ee
where $F(\omega,\xi)$ is the Fourier transform of $f$. Now
\[
p^{q+4}=\left[-\frac{1}{\left(q+1\right)\xi}\right]^{(q+4)/(q+1)}
\]
so that Eq.~\eqref{eq:momdif4} can be written
\[
-\frac{i\omega}{D_0}\left(-\frac{1}{q+1}\right)^{(q+4)/(q+1)}\xi^{-(q+4)/(q+1)}F=\pd[^2F]{\xi^2}.
\]

With $\alpha=-i\omega/D_0[-(q+1)]^{-(q+4)/(q+1)}$ and $m=(q+4)/(q+1)$ one has
\be
\pd[^2F]{\xi^2}+\alpha\xi^{-m}F=0,
\ee
which has the solution
\be
F(\omega,\xi)=F_0(\omega)\sqrt\xi\;J_{1/(2-m)}\left(\frac{2\sqrt\alpha}{2-m}\,\xi^{(2-m)/2}\right),
\ee
where $J_n(z)$ is the Bessel function of the first kind of order $n$ and where $F_0(\omega)$ is constant in $\xi$. Then
\be
f(t,\xi)=\sqrt\xi\uint\df\omega\;F_0(\omega)\text e^{-i\omega t}J_{1/(2-m)}\left(\frac{2\sqrt\alpha}{2-m}\:\xi^{(2-m)/2}\right).
\ee
We note that $2-m=(q-2)/(q+1)$.

Suppose that $F_0(\omega)$ is a simple power in $\omega$, i.e., $F_0(\omega)=f_0(-i\omega)^\gamma$. Now set $\alpha=\alpha_0(-i\omega)$ so that
\begin{align}
f(t,\xi)&=f_0\sqrt\xi\uint\df\omega\,\left(-i\omega\right)^\gamma\text e^{-(i\omega)t}\nonumber\\
&\times J_{1/(2-m)}\left(\frac{2\sqrt{\alpha_0}}{2-m}\sqrt{\xi^{2-m}\left(-i\omega\right)}\right).
\end{align}
Set $-i\omega\xi^{2-m}\equiv\sigma^2$ so that $2\sigma\,\df\sigma=-i\df\omega\,\xi^{2-m}$ and $\sigma=\pm\xi^{(2-m)/2}\text e^{3i\pi/4}\omega^{1/2}$. Careful inspection of the integration limits shows that
\bs\label{eq:Cpm}
\begin{align}
\frac{\sigma}{\sqrt\omega}\xrightarrow{\omega\to-\infty}\pm\xi^{(2-m)/2}\text e^{i\pi/4}&\equiv E_1\\
\frac{\sigma}{\sqrt\omega}\xrightarrow{\omega\to+\infty}\pm\xi^{(2-m)/2}\text e^{3i\pi/4}&\equiv E_2
\end{align}
\es
The resulting inverse Fourier integral of the distribution function then reads
\begin{align}
f(t,\xi)&=2if_0\xi^{(2\gamma+1)(m-2)}\int_{E_1}^{E_2}\df\sigma\;\sigma^{2\gamma+1}\nonumber\\
&\times J_{1/(2-m)}\left(\frac{2\sqrt{\alpha_0}\,\sigma}{2-m}\right)\exp\!\left(-\sigma^2\xi^{m-2}t\right), \label{eq:invF}
\end{align}
which, at $t=0$, has a $\xi$ dependence as $f(0,\xi)\propto\xi^{(2\gamma+1)(m-2)}$. Now originally, $f(0,p)$ is given in the form $f\propto p^{-a}$ and $\xi\propto p^{-(q+1)}$. From Eq.~\eqref{eq:invF} it then follows that
\be
\gamma=-\frac{1}{2}+\frac{a}{2\left(q+1\right)\left(m-2\right)}
\ee
which connects the expression for $f$ with the frequency power-law $F_0(\omega)\propto\omega^\gamma$. And so one has the solution, provided no limits are set on the $p^{-a}$ behavior. Furthermore, for finite times, $t>0$, one then has a relatively simple integral for $f(t,\xi)$ from Eq.~\eqref{eq:invF}. We note also that, if $F_0(\omega)=\sum f_n\omega^n$ then superposition of inverse Fourier integrals of the above type gives the general solution. Thus for any $f(p)$ that is expandible in orthonormal functions expressed as power combinations through a Gram-Schmidt orthonormalization process one has the solution by superposition.

\section{Non-linear momentum diffusion}\label{app:momdiff}
\setcounter{figure}{0}
\renewcommand{\thefigure}{\thesection\arabic{figure}}

\begin{figure}[tb]
\centering
\includegraphics[width=\linewidth]{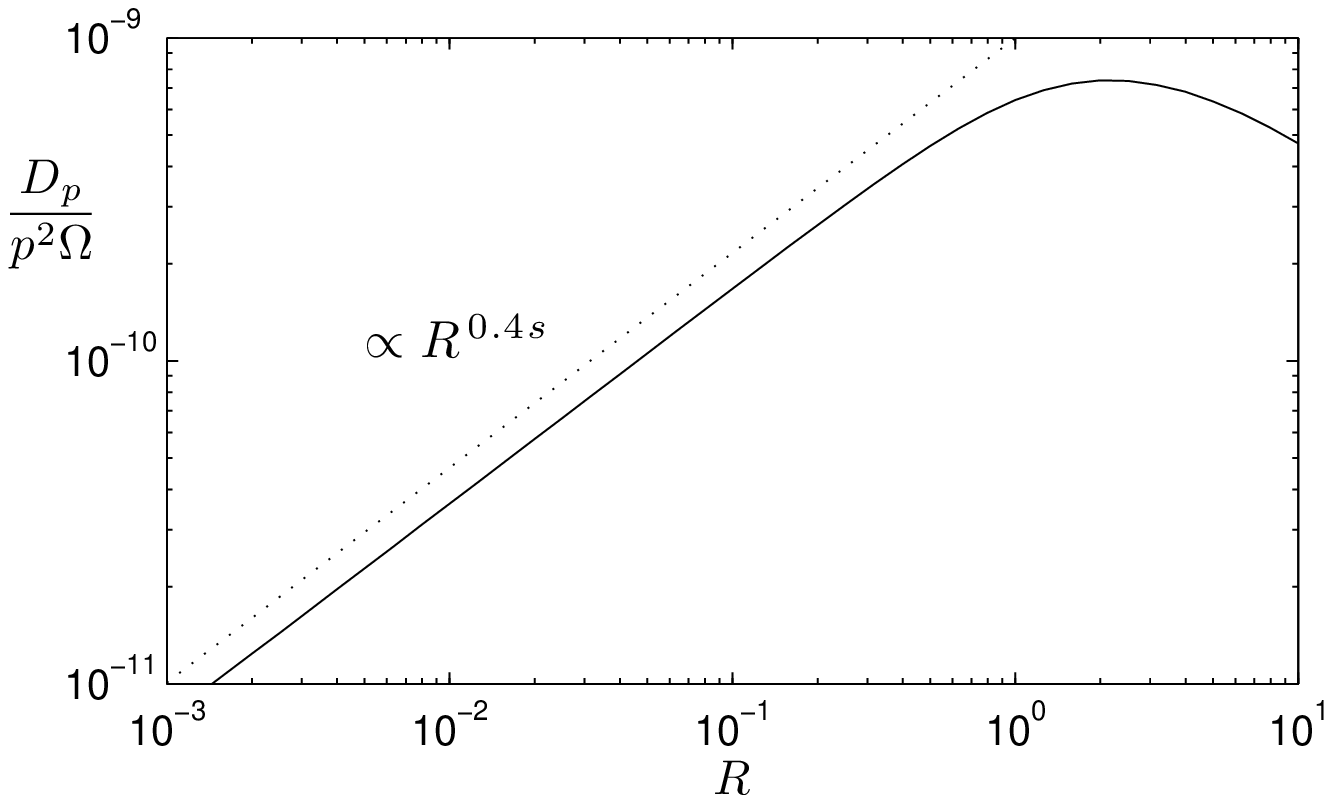}
\caption{Quasi-linear result for the momentum diffusion coefficient (solid line). The dotted line is the power-law fit, which yields $D_p\propto p^{0.4s}$.}
\label{ab:DpFit}
\end{figure}

With $\eps\equiv\rA/R$, the general expression for the Fokker-Planck coefficient of momentum diffusion is of the form
\begin{align}
D_{p}&=\frac{\pi\Om^2p^2\eps^2}{2B_0^2}\int_{-1}^1\df\mu\;\left(1-\mu^2\right)\uint\df k\pa\;G(k\pa)\nonumber\\
&\times\sum_{j=\pm1}\left\{\delta\left[\left(v\mu-j\vA\right)k\pa+\Om\right]+\delta\left[\left(v\mu-j\vA\right)k\pa-\Om\right]\right\},
\end{align}
which was derived under the assumption of slab linearly polarized, undamped Alfv\'en waves \citep{rs:rays}.

Using quasi-linear theory \citep{jok66:qlt}, the momentum diffusion coefficient for Alfv\'en turbulence with vanishing cross helicity is given through \citep{rs:rays}
\be\label{eq:DpQLT}
\frac{D_p}{p^2\Om}=\frac{\eps^2}{R}\left(\frac{\delta B}{B_0}\right)^2C(s)\int_0^1\df\mu\sum_{j=\pm1}\frac{1-\mu^2}{\abs{\mu-j\eps}}\;\tilde G\!\left(\frac{1}{R}\,\frac{1}{\abs{\mu-j\eps}}\right),
\ee
where
\bs
\be
C(s)=\frac{1}{2\sqrt{\pi}}\,\frac{\Ga(s/2)}{\Ga\bigl((1-s)/2)}
\ee
ensures the normalization of the spectrum from Eq.~\eqref{eq:spect}, from which only the wavenumber-dependent part
\be
\tilde G(k)=\left[1+\left(\ell_0k\right)^2\right]^{-s/2}
\ee
\es
is used here. For the spectral index, $s=5/3$ is chosen in accordance with Eq.~\eqref{eq:spect}. In the limit $R\ll1$, the resulting momentum diffusion coefficient can be fitted by a power law $D_p\propto p^{0.4s}$ (see Fig.~\ref{ab:DpFit}).

To obtain an average momentum diffusion coefficient for a distribution function from Eq.~\eqref{eq:dist}, an expectation value has to be calculated as
\be
\m{D_p}=\int_{R_{\text{min}}}^{R_{\text{max}}}\df R\;D_p(R)f(r,A).
\ee

For the parameters given in Table~\ref{ta:param}, one has
\be
\m{D_p}=
\begin{cases}
2.2\times10^{-6}, & \text{for }\rA=10^{-4}\\
3.3\times10^{-4}, & \text{for }\rA=10^{-3}
\end{cases},
\ee
which is in approximate agreement with the simulation results shown in Fig.~\ref{ab:TmomMH}.

A recent investigation of \citet{sha06:mom} using the weakly non-linear theory \citep{sha04:wnl} showed that, compared to previous, quasi-linear results \citep{sha04:mhd}, the velocity-dependence of the momentum diffusion coefficients is flattened. Here, the second-order quasi-linear theory \citep[SOQLT, see][]{sha05:soq,tau08:soq,tau10:soq} is used to derive the non-linear equivalent to judge the resulting differences.

\begin{figure}[tb]
\centering
\includegraphics[width=0.95\linewidth]{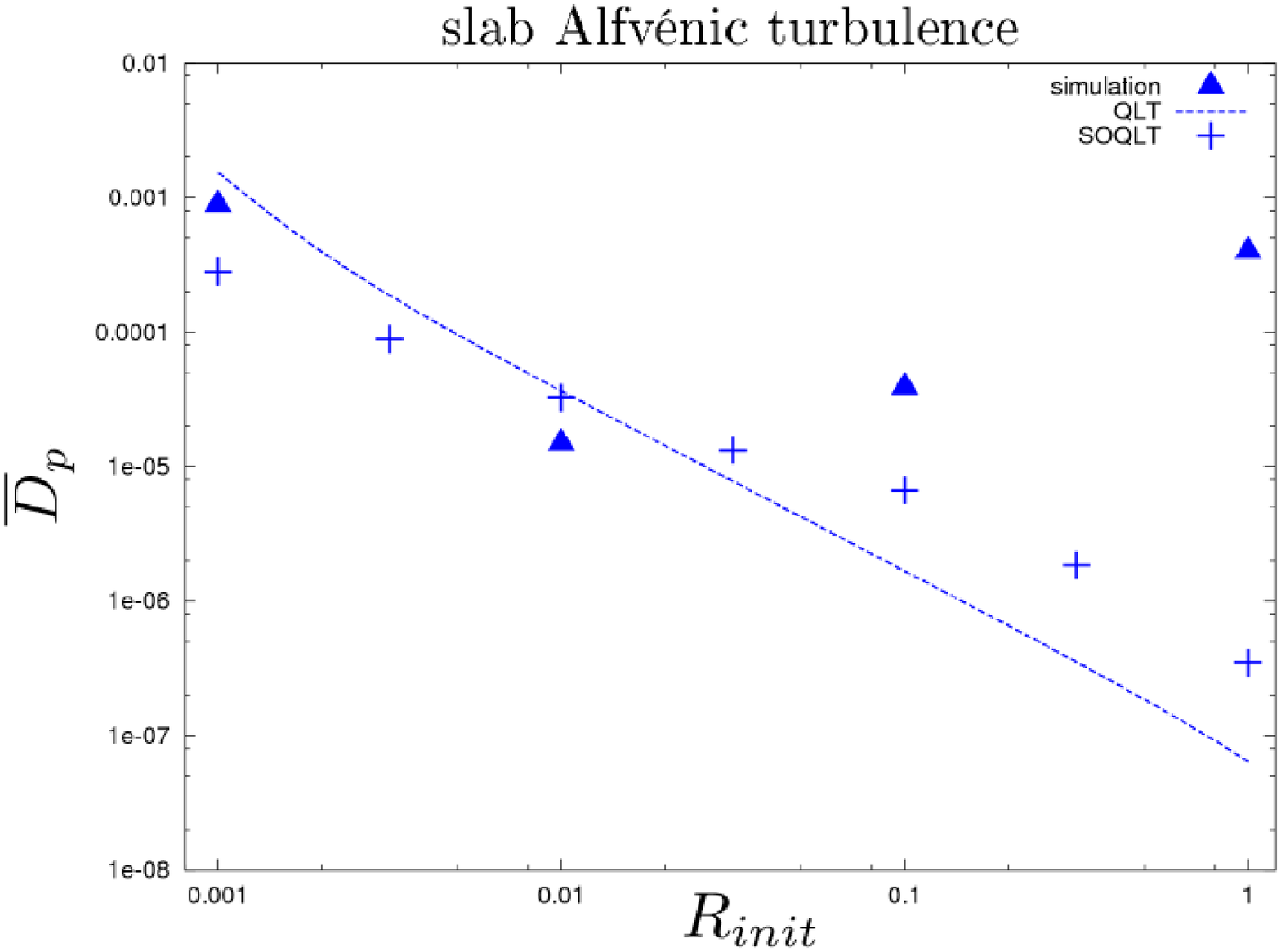}
\caption{Comparison of the momentum diffusion coefficient $D_p$ as obtained from numerical test-particle simulations (triangles), quasi-linear theory (dashed line), and second-order quasi-linear theory (crosses).}
\label{ab:DpSOQLT}
\end{figure}

While in QLT, particles are assumed to follow the unperturbed spiral trajectory, SOQLT takes into account the distribution of particles around their unperturbed orbits through resonance broadening. For the Fourier transform, the new approach involves deviations from the unperturbed orbit, $\bar{\f x}(t)$, as
\be
\m{\text e^{i\f k\cdot\f x}}=
\begin{cases}
{\displaystyle\;\text e^{i\f k\cdot\bar{\f x}(t)}}, & \text{QLT}\\[2pt]
{\displaystyle\;\uint\df z\;f(z)\text e^{ikz}}, & \text{SOQLT},
\end{cases}
\ee
where a Gaussian shape is assumed for the distribution function so that
\be
f(z,t)=\frac{1}{\sqrt{2\pi\sigma(t)}}\,\exp\!\left[\frac{\left(z-\m{z(t)}\right)^2}{2\sigma_z^2(t)}\right]
\ee
with $\m{z(t)}=\m{\bar z(t)}=0$ the (vanishing) mean of the unperturbed parallel coordinate.

The mean square parallel coordinate $\langle z^2(t)\rangle$ that enters the width of the Gaussian as $\sigma_z^2(t)=\langle z^2(t)\rangle-\langle z(t)\rangle^2$ is obtained using QLT, thereby justifying the designation of the resulting theory as being of the second order. In the limit of large times $t\gg\Om^{-1}$ and for pitch-angles close to $90^\circ$, simple analytical expressions can be obtained for $\langle z^2(t)\rangle$, thereby modifying the resonance function. This has a deep impact on some transport parameters. For instance, the mean free path in isotropic magnetostatic turbulence \citep{tau06:sta,tau08:soq} is infinitely large in QLT, but in SOQLT agrees well with simulations \citep{tau12:nov}.

For the momentum-diffusion coefficient, the result reads
\bs
\begin{align}
\frac{D_p}{p^2\Om}&=\frac{\eps^2}{4R}\left(1-\mu^2\right)\left(\frac{\delta B}{B_0}\right)^2C(s)\int_{-1}^1\df\mu\;\int_0^\infty\df x\;\tilde G(x)\nonumber\\
&\times\sum_{j=\pm1}\left(\mathcal R_++\mathcal R_-\right), \label{eq:DpSOQ}
\end{align}
where
\be
\mathcal R_\pm=x\sqrt\pi\,\frac{B_0}{\delta B}\exp\left[-x^2\left(\frac{B_0}{\delta B}\right)^2\bigl((\mu-j\eps)x\pm R^{-1}\bigr)^2\right]
\ee
\es
is the second-order resonance function. The integrals in Eq.~\eqref{eq:DpSOQ} cannot be solved analytically so that a numerical solution, which in this case is intricate, has to be evoked.

The result for $D_p$ is shown in Fig.~\ref{ab:DpSOQLT} in comparison to results obtained from a test-particle simulation and the quasi-linear result from Eq.~\eqref{eq:DpQLT}. At first view, the SOQLT result seems to be an improvement compared to QLT for rigidities larger than $R=0.0015$ but one has to be careful before coming to any conclusions. A conspicuous feature of the values obtained with SOQLT can be seen at the lower rigidity range. The values exhibit a discrepancy in comparison to the simulation, because the slope of the curve is positive for rigidities smaller than $R=0.0015$. As was pointed out in the preceding, the simulation in this parameter regime provides the most reliable results, since for small rigidities it reached almost diffusive behavior. Therefore, one would expect a decreasing slope everywhere and the values for the Fokker-Planck coefficient to be considerably higher in this range. In the calculation, the Alfv\'en rigidity is set to $\rA=10^{-3}$ in the calculation, hence the approximation is no longer valid in this regime.

There is an additional feature of the simulations that has to be taken into account. As shown in Fig.~\ref{ab:TmomMH}, the momentum diffusion coefficient decreases for a longer time if the ratio $\eps=\rA/R$ is decreased, i.e., for larger initial particle rigidities. Therefore, longer simulation times would be required in principle to ensure that the diffusive regime is reached for all initial rigidities. Therefore, the values of the momentum diffusion coefficient are overestimated for larger initial rigidities. However, quantitative statements about the expected reduction are not possible; instead, simulations with longer running times would be required, which are beyond what is currently available.



\end{document}